\documentclass[runningheads]{llncs} 
\usepackage{hyperref}
\usepackage{graphicx}
\usepackage{caption} % required for subfigures
\usepackage{subcaption} % required for the temporary ITA definition
\usepackage{enumitem} % required for links to challenges
\usepackage{amsmath} % required for the \text command

\begin{document}
%
%\title{Contribution Title\thanks{Supported by organization x.}}
% \title{Fairness Across Skin Tones in Deep Learning Dermatology: Is the Problem Solved?}
%
% \titlerunning{Fairness Across Skin Tones in Deep Learning Dermatology}

% \title{Revisiting skin tone fairness in dermatology}
\title{Revisiting Skin Tone Fairness in Dermatological Lesion Classification}
\titlerunning{Revisiting Skin Tone Fairness in Dermatology}
\author{Thorsten Kalb\inst{1}%\orcidID{0009-0004-0561-3479}%index{Kalb, Thorsten}
\and
Kaisar Kushibar\inst{1}%\orcidID{0000-0001-7507-5208} %index{Kushibar, Kaisar}
\and
Celia Cintas\inst{2}%\orcidID{0000-0002-8064-9189} %index{Cintas, Celia}
\and
Karim Lekadir\inst{1}%\orcidID{0000-0002-9456-1612} %index{Lekadir, Karim}
\and
Oliver Diaz\inst{1}%\orcidID{0000-0001-6789-5177} %index{Diaz, Oliver}
\and
Richard Osuala\inst{1}%\orcidID{0000-0003-1835-8564} %index{Osuala, Richard}
}
\authorrunning{T. Kalb et al.}
\institute{{%Barcelona Artificial Intelligence in Medicine Lab (BCN-AIM), 
Departament de Matemàtiques i Informàtica, Universitat de Barcelona, %08007 Barcelona, 
Spain} \\
%\email{thkalbka98@alumnes.ub.edu}
%\email{\{kaisar.kushibar,karim.lekadir,oliver.diaz,richard.osuala\}@ub.edu}\\
\email{richard.osuala@ub.edu}
\and
IBM Research Africa, Nairobi, Kenya \\
%\email{celia.cintas@ibm.com}\\
}
\maketitle              % typeset the header of the contribution
\begin{abstract}

%Recent advancements in deep learning methods have shown remarkable promise for enhancing skin lesion diagnosis.
%Sentence 1: The promise of deep learning for improved skin cancer diagnosis.
%With deep learning models in dermatology approaching a state of maturity that advocates for clinical deployments, it becomes increasingly important to investigate and mitigate model biases towards different subpopulations and pigmentation types, given that skin diseases appear differently across skin tones. Biased models and datasets could contribute to and amplify disparities in skin disease diagnosis.  %such as ethnic, socio-economic and demographic groups or pigmentation types. 
%Sentence 2: The issue that desirable fairness characteristics of these algorithms are at best questionable
%Particularly questionable is classification fairness across skin tones where dark lesions can contrast less with surrounding healthy dark skin.
%Sentence 3: Example of fairness across skin types.
%With the absence of reliable skin tone labels in lesion patch datasets, the characterisation of skin tones is challenging, which restrains fairness assessments.

%In dermatology, as deep learning models mature and move towards clinical deployment, addressing biases towards various subpopulations and pigmentation types becomes crucial.
Addressing fairness in lesion classification from dermatological images is crucial due to variations in how skin diseases manifest across skin tones. 
%Biased models and datasets could amplify disparities in skin disease diagnosis.
However, the absence of skin tone labels in public datasets hinders building a fair classifier. To date, such skin tone labels have been estimated prior to fairness analysis in independent studies using the Individual Typology Angle (ITA).
%Sentence 4: The necessity of estimating skin types (e.g. based on ITA) to be able to do a fairness across skin types study.
% In this work, we provide a comparative analysis of different ITA based strategies to automatically estimate skin tones.
Briefly, ITA calculates an angle based on pixels extracted from skin images taking into account the lightness and yellow-blue tints. These angles are then categorised into skin tones that are subsequently used to analyse fairness in skin cancer classification. %algorithms. #
In this work, we review and compare four ITA-based approaches of skin tone classification on the ISIC18 dataset, a common benchmark for assessing %the fairness of skin cancer classification methods 
skin cancer classification fairness in the literature. Our analyses reveal a high disagreement among previously published studies demonstrating the risks of ITA-based skin tone estimation methods. Moreover, we investigate the causes of such large discrepancy among these approaches and find that the lack of diversity in the ISIC18 dataset limits its use as a testbed for fairness analysis.
Finally, we recommend further research on robust ITA estimation and diverse dataset acquisition with skin tone annotation to facilitate %reproducible and 
conclusive fairness assessments of artificial intelligence tools in dermatology.
Our code is available at \href{https://github.com/tkalbl/RevisitingSkinToneFairness}{https://github.com/tkalbl/RevisitingSkinToneFairness}.
%The abstract should briefly summarize the contents of the paper in 150--250 words.

\end{abstract}

\section{Introduction}
Skin cancer is one of the most prevalent cancer types worldwide \cite{WHO}. According to~\cite{cancerorg}, early detection increases the survival rate to 99$\%$ compared to 32$\%$ in late stage detection.
%The International Skin Imaging Collaboration (ISIC) fosters challenges and image repositories to help the scientific community to advance in the development of Artificial Intelligence (AI) models to support in this detection task \cite{ISIC18Comparison}. 
%This study introduces the problem of robustness in dermatology classification i.e. 'a minor dermoscopic image changes suffice to influence the Convolutional Neural Networks (CNNs)' diagnosis.' \href{https://www.sciencedirect.com/science/article/pii/S0959804920313575}{paper}. Not sure how this is related. \\
%However, it is important to ensure that these models do not discriminate individuals.
%Discrimination by ethnicity or skin type is especially important, as 
The 5-year survival rates after surgical removal of melanoma have been shown to be lower for black patients (73$\%$) compared to white ($88\%$), %or patients from other ethnic groups (85$\%$), 
although melanoma is 23 times more prevalent in white patients %than in black patients 
~\cite{SurvivalMELRace}. 
%While reasons for such disparities remain poorly understood \cite{SurvivalMELRace}, it can be assumed due to contrast and with other things being equal, to be a more difficult task to detect dark lesions (e.g., melanoma) in patients of darker skin tones compared to lighter skin tones. 
% Lam et al.~\cite{lam2022racial} showed that reduced survival rates for ethnic minority patients compared to Caucasian patients may be attributed to several contributing factors. First, a longer time to diagnosis and, as more advanced stages, can lead to a poorer prognosis~\cite{dawes2016racial,hu2009disparity,hu2006comparison}.
Dick et al.~\cite{dick2019odds} found that black patients were significantly more likely to present with advanced-stage disease, even after adjusting for tumour characteristics and demographic factors. 
%Second, socioeconomic status differences may lead to increased barriers that limit access to medical care in minority populations. Communities of lower socioeconomic status tend to have a decreased density of dermatologists, further increasing the disparity of access to care~\cite{vaidya2018socioeconomic}.
Deep learning models have demonstrated a remarkable performance in skin lesion analysis and classification~\cite{esteva2017dermatologist}. Therefore, they are promising tools to detect skin cancer earlier and, thus, in theory, bear the potential to reduce the aforementioned disparities.
In practice, however, deep learning models have been shown to be prone to and exacerbate existing societal biases~\cite{EthicalAI,FairPrune2022,birhane2023hate}.
Although bias and fairness assessment in skin lesion classification has been an active research area \cite{XiaoXiao,FairPrune2022,BevanArtifacts}, there is a substantial limitation on developing an unbiased classifier. That is, many publicly available datasets lack information about ethnicity or skin types~\cite{SkinTypeDiversity}. %Approximately two-thirds of dermatological datasets lack metadata regarding ethnicity, and only approximately $6\%$ report a dermatological skin type~\cite{SkinTypeDiversity}.
Hence, such labels are usually obtained using different automated methods.
%and used as ground truths for further studies and evaluations.
One of the common approaches is based on Individual Topology Angle (ITA) (see Section~\ref{sec:AutoITAMethods}) that have been used in several studies~\cite{SkinToneBevan2022,derm_ita,Groh,IBM,Loaiza,XiaoXiao}. For instance, Kinyanjui et al. \cite{IBM} estimated skin tones to assess their effect on lesion classification performance, while Bevan et al.~\cite{SkinToneBevan2022} labelled skin tones for model debiasing. These previous works~\cite{IBM,Loaiza,SkinToneBevan2022,XiaoXiao} form the basis of our analysis and are further described in detail in Section~\ref{sec:AutoITAMethods}.
% For example, in \cite{SkinToneBevan2022}, the authors 

These existing studies utilise ITA-based estimation of skin tones as a proxy to ground truth. However, in this work, we show that there is a large disagreement in the assigned skin tones on the same dataset, which suggests that the reported results may be inconclusive.
Therefore, %, in this paper, 
we investigate the causes and extent of the discrepancies between the estimated skin tones in previous studies. We uncover the complexities, pitfalls, and differences across ITA-based skin tone estimation techniques that question the conclusions derived in previous studies. In summary, our contributions are as follows:
\begin{itemize}
    \item We compare different ITA-based automatic skin tone estimation algorithms and highlight  common pitfalls.
    \item We demonstrate the impact of different skin tone estimations on the outcome of fairness analyses of the same model.
    \item We show the limitations of the commonly used ISIC18 dataset \cite{ISIC18} to assess fairness of skin lesion classifiers.
\end{itemize}

\section{Methods and Materials}

\subsection{Dataset}\label{sec:ImageDataset}
The ISIC18 dataset \cite{ISIC18} includes 10015 dermatoscopic images from Austria and Australia.
% The ISIC2018 (ISIC18) dataset includes dermatoscopic images from two different populations acquired in two clinical centres -- Austria and Australia -- over a period of 20 years. 
%The dataset is categorised into image segmentation (Task 1), clinical feature detection (Task 2), and disease classification (Task 3). 
The present work focuses on classification fairness of seven skin lesion types distributed as:
% data comprising 10015 dermoscopic images of seven skin lesion types confirmed either via pathology (53.3$\%$), follow-up, expert consensus, or in-vivo confocal microscopy.
% These include 
1113 melanoma (MEL), 6705 melanocytic nevus (NV), 514 basal cell carcinoma (BCC), 327 actinic keratosis (AKIEC), 1099 benign keratosis (BKL), 115 dermatofibroma (DF), and 142 vascular lesions (VASC).

\subsection{Evaluation of skin lesion classification}%\label{sec:ImageDataset}
To build a skin lesion classification model, we used a %lightweight 
MobileNetV2 \cite{mobilenetv2} %architecture 
initialised with ImageNet weights \cite{ImageNet}. Only the last layer was replaced to correspond with the seven skin lesion classes. %The choice of architecture was based on the universality principle proposed in the FUTURE-AI guidelines \cite{EthicalAI}. Thus, facilitating universal applicability including in low-resource settings.
The choice of this model was driven by its universal applicability \cite{EthicalAI} due to its lightweight architecture that facilitates deployment across a wide range of devices including portable devices in low-resource %healthcare 
settings.
% MobileNetV2 \cite{mobilenetv2} pretrained on ImageNet \cite{ImageNet} was chosen to evaluate skin lesion classification in line with the FUTURE-AI universality principle \cite{EthicalAI}: MobileNetV2 is a 53 layer convolutional neural network and, as lightweight model, it allows for deployment across diverse dermatological devices, hencewith facilitating universal applicability across health care environments including low-resource settings.
%The MobileNetV2 was further pretrained on ImageNet \cite{ImageNet} and implemented in Keras. %\cite{Keras}.
% The last layer of the pretrained MobileNetV2 is replaced with seven softmax neurons corresponding to the seven skin lesion classes.
%The learning rate and batch size were optimised with a grid search, 
A grid search was performed for a combination of (a) a constant learning rate ranging from 1e-3 to 1e-6 in steps of 10 on a logarithmic scale and (b) the batch size, where 12, 24, 32, 48, and 64 were tested. 
Based on validation loss, the grid search yielded a learning rate of 1e-5 and a batch size of 16. The network was optimised using Adam to minimise the sparse categorical cross entropy loss for a maximum of 60 epochs with an early stopping policy to terminate training if the validation loss did not improve for 30 epochs. All experiments were performed on an NVIDIA GeForce RTX 2080 SUPER with 8GiB memory.
% A grid search was performed yielding a learning rate of 1e-5 and a batch size of 16. All layers are set to trainable and no data-augmentation was applied for the 600×450 pixel input images. The models are trained for 60 epochs on a NVIDIA GeForce RTX 2080 SUPER with 8192 MiB memory. Accounting for class imbalance, a weighted sparse categorical cross entropy loss is optimised using an Adam optimiser that outperformed SGD in preliminary experiments. %, which seemed to outperform stochastic gradient descent. 
% Early stopping is applied whenever the validation loss did not improve for more than 30 epochs, which commonly occurred after 14 epochs. The model from the best epoch in terms of validation loss is saved and used for predictions on the test set.
All evaluations were conducted three times with different seeds for splitting into training (57$\%$), validation (14$\%$) and test (29$\%$) sets, stratified by lesion, with a slightly different split for the RP data shift experiment (see Section~\ref{sec:DataShiftExperiments}).
%an exception being the data shift experiments (see Section~\ref{sec:DataShiftExperiments}).
% All evalutations were conducted three times with different seeds for splitting into training, validation and test sets, stratified by lesion.  
%(see Table~\ref{tab:lesions}).
%The evaluation was based on [insert metrics]...
 
\subsection{Skin tone estimation}
\label{sec:AutoITAMethods}

\subsubsection{Individual Topology Angle (ITA)}
\begin{figure}[tb]
\centering
\scalebox{0.86}{
\begin{subfigure}[c]{0.4\textwidth}
         \centering
         \includegraphics[width=\textwidth]{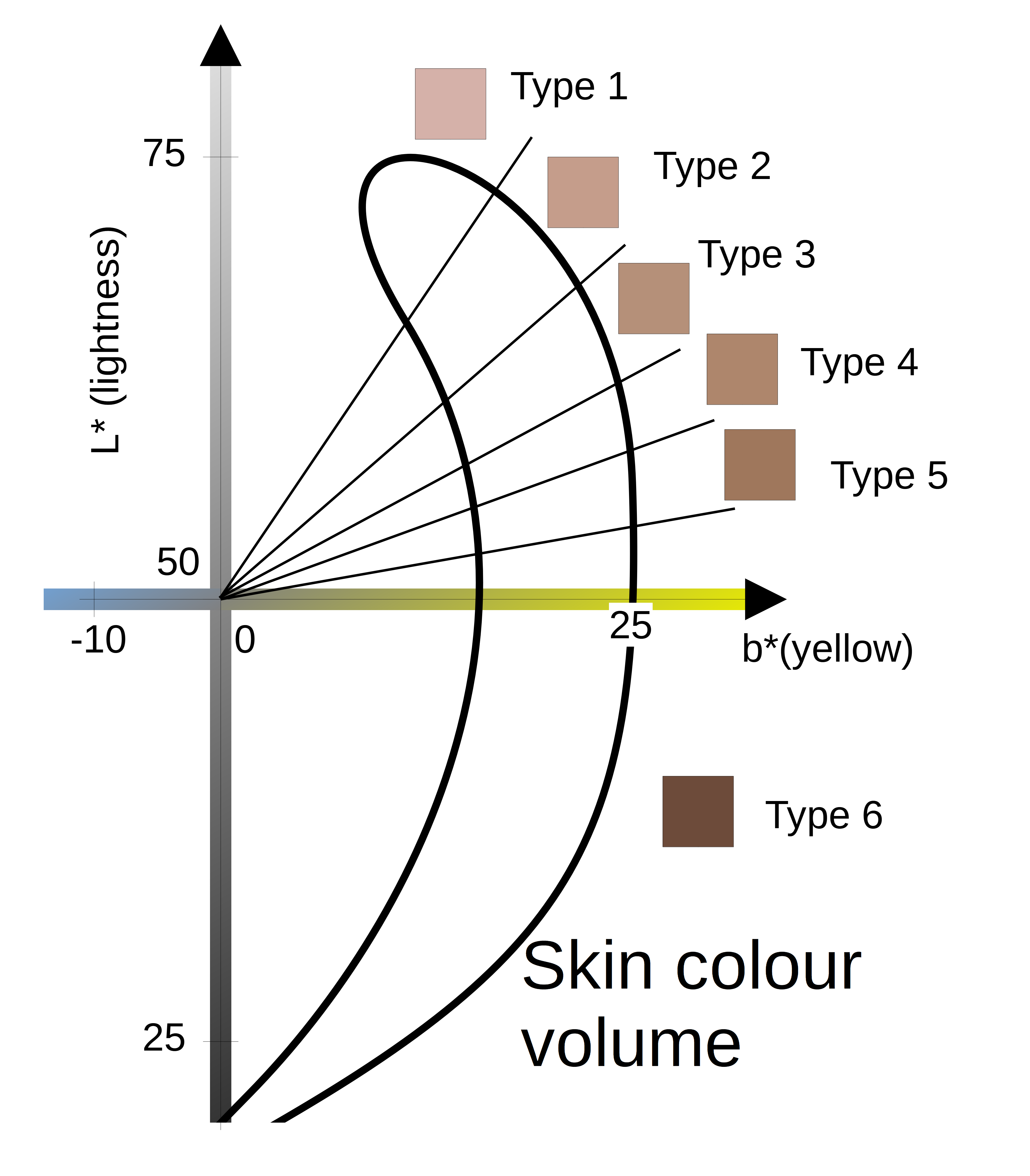}
\subcaption{\label{subfig:ITA}}
\end{subfigure}
}
\begin{subfigure}[c]{0.55\textwidth}
\centering
\begin{tabular}{l r }
 \textbf{Skin Type} & \textbf{ITA range}\\
Type 1 & $\text{ITA}> 55 ^\circ $ \\
Type 2 & $ 55 ^\circ \geq \text{ITA} > 41 ^\circ $ \\
Type 3 & $ 41 ^\circ \geq \text{ITA} > 28 ^\circ $ \\
Type 4 & $ 28 ^\circ \geq \text{ITA} > 19 ^\circ $ \\
Type 5 & $ 19 ^\circ \geq \text{ITA} > 10 ^\circ $ \\
Type 6 & $ 10 ^\circ \geq \text{ITA} \phantom{{} > 10 ^\circ }$ \\
\end{tabular}
\subcaption{\label{subfig:ITAthresholds}}
\end{subfigure}
\caption{(a) Skin colour volume in the $L*$ - $b*$ plane of CIELab colour space with ITA thresholds \cite{SkinToneBevan2022}.
% Intuitively, the ITA can be interpreted as a pointer on the skin colour volume.
(b) Skin types from ITA thresholds, as in \cite{SkinToneBevan2022}.
% These thresholds and a representative skin colour of the corresponding ITA bins are shown in \ref{subfig:ITA}
}
\label{tab:ITABevan}
\end{figure}

In the absence of dermatologist-confirmed skin type labels, researchers proposed automatic skin tone and skin type estimations \cite{IBM,Loaiza,Groh,derm_ita,XiaoXiao,SkinToneBevan2022}, which are commonly based on the ITA \cite{ITA1991}. The ITA is defined within the L*-b*-plane of the CIELab colour space, according to Equation \ref{eqn:ITA}, and illustrated in Figure \ref{subfig:ITA}.
\begin{equation}
\text{ITA} = \arctan \left( \frac{L-50}{b} \right) \cdot \frac{180^{\circ}}{\pi}
\label{eqn:ITA}
\end{equation}
ITA values from different studies only become comparable when the lighting conditions and measurement devices are known and corrected for \cite{Colorimetry}. Categorical skin types are obtained by binning ITA values \cite{IBM,Groh,Colorimetry} as shown in Figure \ref{subfig:ITAthresholds}. 
%\footnote{
We note that there is no consensus for ITA binning thresholds and high uncertainty for any ITA (colour) to Fitzpatrick \cite{FitzpatrickVI} (sun reactivity) mapping.
%}
% (skin colour can change after sun exposure).
%
%
%\begin{figure}
%\centering
%\includegraphics[width=0.5\textwidth]{Figures/ITA_def.png}
%\caption{Illustration of the Individual Topology Angle (ITA) in the L*-b*-plane of the CIELab colour space. The black curved shape shows the skin colour volume. ITA thresholds from \cite{SkinToneBevan2022} are shown as black lines together with a representative skin colour of resulting ITA bins (skin types). Intuitively, the ITA can be interpreted as a pointer on the skin colour volume.} \label{fig:ITA_def}
%\end{figure}
%
Given our analysis of four existing ITA-based automatic skin tone estimation methods \cite{IBM,Loaiza,SkinToneBevan2022,XiaoXiao}, we note the following key issues that such methods need to address.
%Before analysing four existing ITA-based automatic skin tone estimation methods\cite{IBM,Loaiza,SkinToneBevan2022,XiaoXiao}, we note the following key issues that such methods face.
%
\begin{enumerate}[label=\textbf{I.\arabic*},ref=I.\arabic*]
    \item%[I1] 
    \label{ch:preprocessing} \textit{Lighting conditions}: The ITA is sensitive to illumination, especially to brightness or lack of yellow chroma.
    \item%[I2] 
    \label{ch:skin} \textit{Non-skin imaging artefacts}: The ITA is defined for any colour, but only meaningful for skin. Estimating ITA, in part, for hair, artefacts and dark borders can create misleading results.
    \item%[I3] 
    \label{ch:lesion} \textit{Lesion to skin contrast}: The pigmented lesion is not representative for the skin colour of the patient and needs to be excluded.
    \item%[I4] 
    \label{ch:ITA} \textit{From pixel to image-level}: The ITA is defined for each pixel. As one representative ITA value needs to be assigned to an image, there is no consensus how to address variance and outliers of the ITA distribution.
\end{enumerate}

\noindent In the following, we describe the aforementioned four existing skin tone estimation methods \cite{IBM,Loaiza,SkinToneBevan2022,XiaoXiao}, which are empirically analysed in Section \ref{sec:experiments}.

\subsubsection{Method 1: Deep learning-based skin segmentation}
% We adopt the deep learning-based healthy skin segmentation (DLHSS) proposed in \cite{IBM} using a Mask R-CNN \cite{MaskRCNN} trained on the negative of the lesion segmentation masks from ISIC18 \cite{ISIC2018}.
% The model was then fine-tuned on 343 healthy skin images manually segmented by \cite{IBM} from the SD-198 dataset \cite{SD_dataset} to address issues \ref{ch:skin} and \ref{ch:lesion}.
We adopt a skin segmentation model kindly provided by the authors of \cite{IBM}, which is a Mask R-CNN trained with manually segmented ISIC18 images to address \ref{ch:skin} and \ref{ch:lesion}.
The segmented pixels of healthy skin are converted to CIELab colour space, for which the median within one standard deviation of the mean of ITA's $L*$ and $b*$ components is calculated. Based on these $L*$ and $b*$ median values, an image's ITA is calculated according to Equation \ref{eqn:ITA}. Selecting $L*$ and $b*$ median values separately can help to avoid outliers, but may also lead to less precise ITAs, as $L*$ and $b*$ medians likely do not correspond to the same pixels.
%
%Note that selecting the median within a standard deviation of the mean and the ITA calculation are not commutative, as the selected $L*$ and $b*$ values might not correspond to the same pixels.
%Kinyanjui et al. \cite{IBM} did not report any pre-processing of the images.
%Their five undefined ITA estimations presumably stem from the Mask R-CNN, when it did not detect any healthy skin instance with a sufficiently high threshold. They provided their ITA labels and a re-trained Mask R-CNN on request.
%For a better comparison with their publication, the original ITA labels are used rather than the re-trained model's predictions.
%
\subsubsection{Method 2: Colour-based skin segmentation}
% As further skin alternative, we adopt the colour-based skin segmentation algorithm proposed by \cite{Loaiza}.  %based on the expected colours of skin
This method follows the skin segmentation algorithm proposed in \cite{Loaiza}.
Input images are converted to grayscale before applying Otsu binarisation and thresholding \cite{otsu1979threshold} to detect pixels that are non-lesion. For original values of healthy skin pixels, different thresholds in HSV and YCrCb colour spaces are applied to define potential skin colours. For the pixels within these thresholds, the mean values for red, green and blue are computed and define a representative skin colour. This skin colour is then converted into CIELab space before applying Equation \ref{eqn:ITA} to calculate the ITA.
%This approach did not take into account that skin lesions might have the colour of darker skin (problem \ref{ch:lesion}), which might explain the comparably large amount of dark skin tone estimations in Figure \ref{fig:SkinToneComparison}.
%Similarly to \cite{IBM}, the undefined ITA labels can be attributed to the skin segmentation, which might not detect any skin.
%Although their code is publicly available, the method will not be compared in this study, as they did not address problems \ref{ch:preprocessing} and \ref{ch:lesion}.

\subsubsection{Method 3: Random patch algorithms}
% Following \cite{SkinToneBevan2022}, we validate a further alternative based on semi-random patches.
Next method is based on semi-random patches proposed in \cite{SkinToneBevan2022}.
It is assumed that a lesion is in the centre of the image and healthy skin is presumably found in at least one of eight patches of $20\times20$ pixels in the periphery.
%Images of different sizes were centre-cropped and resized. 
Before patch extraction, input images are centre-cropped, resized and small dark artefacts such as hair are removed via black-head morphology to address \ref{ch:skin}.
For each patch, the mean ITA is calculated. The ITA of the patch is selected that corresponds to the brightest skin type, which arguably has the effect of having excluded the pigmented lesion.
%With this approach, they argue, the pigmented lesion is likely to be excluded.
%Although they did not pre-process the image in colour space, they applied black-head morphology to remove small dark artefacts such as hair, addressing partially problem \ref{ch:skin}.
%The 38 undefined skin type labels in Figure \ref{fig:SkinToneComparison} are presumably the result of removing duplicates from the dataset and therefore, some images of the ISIC18 dataset \cite{ISIC18} do not appear in their ITA labels.
%Bevan et Atapour-Abarghouei \cite{SkinToneBevan2022} reported an over-classification of Type 6 samples and surmised this might be caused by dark lighting conditions.
%They proposed further research into this phenomenon and published both their skin type labels and their code.
%In this work, we reviewed their code and detected a minor difference compared to \cite{IBM} and \cite{Loaiza}:
In contrast to previous methods, %Bevan et Atapour-Abarghouei 
this method used \textit{arctan} in Equation \ref{eqn:ITA} for the ITA instead of \textit{arctan2}.
%The difference between the two functions is that 
Unlike \textit{arctan}, \textit{arctan2} takes into account the signs of the catheti as described in Equation \ref{eqn:arctan2}.

\begin{equation}
\text{arctan2}(x,y) = \begin{cases}
    \arctan(y/x), & \text{if } x > 0\\
    \arctan(y/x) - \operatorname{sgn} (y)* \pi, & \text{if } x <0\\
    \operatorname{sgn} (y)* \pi & \text{if } x = 0 \\
\end{cases}
\label{eqn:arctan2}
\end{equation}
%
% Using \textit{arctan} instead of \textit{arctan2} in Equation \ref{eqn:ITA} assumes $b*$ cannot be negative.
Using \textit{arctan} in Equation \ref{eqn:ITA} assumes $b*$ cannot be negative.
%This implies that using the arctan instead of the arctan2 for the ITA calculation implicitly assumes that the b* component cannot be negative.
However, a negative $b*$ value encodes blue or absence of yellow chroma.
Although the blue colour in skin is not intuitive, its appearance may depend on variations in illumination, dermatoscope, and camera.
% While it might seem reasonable to assume that skin is not blue, this is not necessarily the case for images of skin, where the colours depend on the illumination, dermatoscope, and the camera.
Therefore, in our analysis, we include both versions of ITA estimation using \textit{arctan} and \textit{arctan2} referred to as RP and RP2, respectively.
% Therefore, we propose an adaption of the Random Patch (RP) algorithm \cite{SkinToneBevan2022}, namely, the Random Patch algorithm based on \textit{arctan2} (RP2). Both RP and RP2 are included in the analysis of this work.

\subsubsection{Method 4: Histogram thresholding with grey-world white balancing}
The next method is adopted from \cite{XiaoXiao}. After centre-cropping and resizing, a grey-world white balance algorithm is applied to correct for light influence addressing \ref{ch:preprocessing}. Next, non-lesion skin is segmented using the Generalized Histogram Threshold \cite{GHT} algorithm.  The segmented area is transformed from RGB to CIELab-space to calculate the ITA.
However, in \cite{XiaoXiao}, reproducibility was limited regarding how one representative ITA is obtained from the multiple segmented pixels.
A further limitation is the assumption that skin images are grey on average.
% A further limitation is the assumption of the grey-world white balance algorithm that images are grey on average, although images of skin presumably have a skin colour on average.

%Although they provided visual examples at each step of their ITA estimation pipeline, they did not publish code or ITA labels.
%Since this algorithm is not reproducible, it will not be included in the further analysis. 
%Noteworthy, from the compared publications, they were the only ones to address problem \ref{ch:preprocessing} and to pre-process the images in colour space.
%% Move this to the discussion part.
%However, a grey-world white balance algorithm assumes that on average the image is grey, although images of skin presumably have a skin colour on average.

\section{Experiments and Results} \label{sec:experiments}

\subsection{Comparison of ITA estimation methods}
\begin{figure}[tb]
\centering
\scalebox{0.94}{
\includegraphics[width=\textwidth]{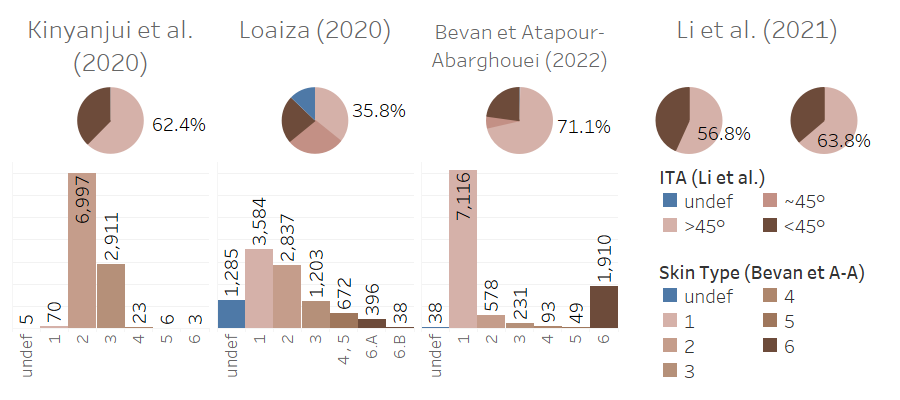}
}
\caption{Comparison of the four different ITA-based skin type estimation methods based on values reported in the literature \cite{IBM,Loaiza,SkinToneBevan2022,XiaoXiao}. Note that Li et al. \cite{XiaoXiao} only reported darker ($ \text{ITA}< 45^{\circ}$, 43.2$\%$) and lighter ($ \text{ITA}> 45^{\circ}$, 63.8$\%$) skin %, compared in the pie charts. 
and that their sum equals %of reported darker and lighter skin sums up to 
107$\%$; it is assumed that one of these values is correct and both possible distributions are shown.} \label{fig:SkinToneComparison}
\end{figure}

\begin{figure}[tp]
\scalebox{0.92}{
\includegraphics[width=\textwidth]{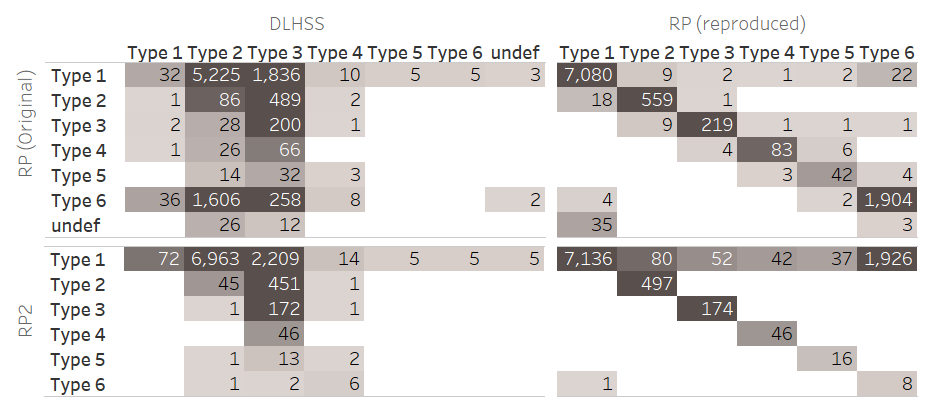}
}
\caption{Comparison of ITA estimates of Deep Learning-based Healthy Skin Segmentation (DLHSS) \cite{IBM}, Random Patch \cite{SkinToneBevan2022} (RP) and Random Patch with \textit{arctan2} (RP2).
% The reproduced ITAs were further compared to published \cite{SkinToneBevan2022} RP ITA labels showing good agreement.
} 
\label{fig:DetailedComparisonITA}
\end{figure}

% Move
%Although their results might differ, all of these three studies \cite{IBM,XiaoXiao,SkinToneBevan2022} proposed an algorithm for automatic skin tone estimation based on the ITA and reported an estimated skin type distribution for the ISIC2018 \cite{ISIC18} dataset. A fourth study \cite{Loaiza} also estimated skin tone distributions of three ISIC datasets, including the ISIC2018 (ISIC18) \cite{ISIC18} dataset. 
A comparison of the estimated skin tone distributions is shown in Figure~\ref{fig:SkinToneComparison}. Although the same thresholds (see Figure \ref{subfig:ITAthresholds}) are applied to all four skin tone estimation methods for ITA binning, their estimates on the ISIC18 dataset clearly differ. 
A detailed comparison of the agreement among the ITA estimation methods is shown in Figure~\ref{fig:DetailedComparisonITA}, where a diagonal matrix would represent perfect agreement.
%
%
% Comparing the original labels between RP and DLHSS there appears to be little agreement on which images show dark skin: Most Type 6 images in RP appear to be Type 2 in DLHSS and most Type 2 images in DLHSS are classified as Type 1 in RP. All Type 5 and 6 images in DLHSS are labelled as Type 1 in RP.
%
There is little agreement between RP and DLHSS, especially for dark skin. Most type 6 images in RP appear as type 2 in DLHSS and most type 2 images in DLHSS are classified as type 1 in RP. All type 5 and 6 images in DLHSS are labelled as type 1 in RP.
Comparing the RP to RP2, the entries of the matrix lay on the diagonal and in the first row. This implies that in most cases, RP and RP2 agree on the same skin type.
However, approximately 21.6$\%$  of the samples are classified as type 1 by RP2 and darker by RP.
Thus, using the \textit{arctan} can account for the over-estimation of dark skin images and affect all except for type 1.
% and not only Type 6 samples are affected, but all skin types except for Type 1 are affected.
%
Comparing DLHSS and RP2, both %estimations 
agree that approximately 99.0$\%$ of the samples %in the ISIC18 dataset \cite{ISIC18} 
show skin at least as light as type 3 ($\text{ITA} > 28^{\circ}$), while only eight out of 10015 samples are classified at least as dark as type 4 ($\text{ITA} \leq 28^{\circ}$). 

\begin{figure}[tb]
\centering
\includegraphics[width=\textwidth,keepaspectratio]{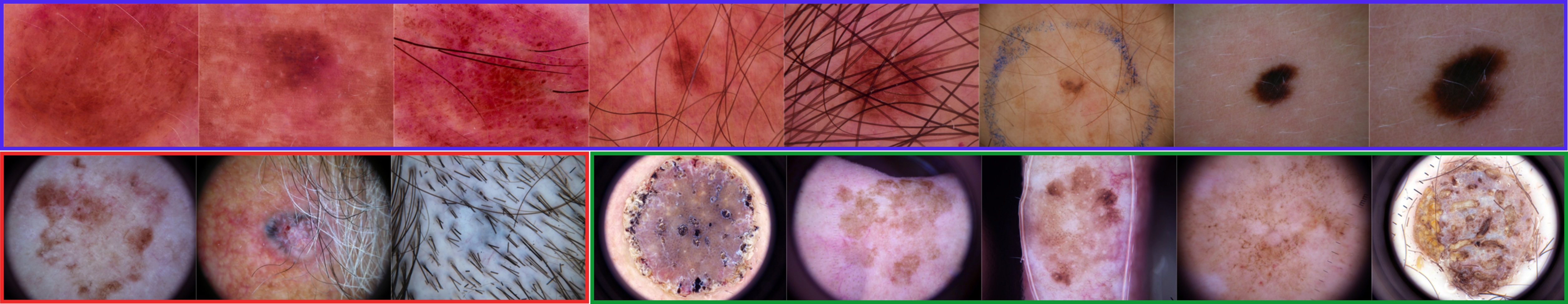}
\caption{Blue and red: All samples where both DLHSS and RP estimate dark skin tones (types 4-6,
$\text{ITA} \leq 28^\circ$). Blue: Dark skin according to DLHSS and RP2. Green: Samples with darkest skin according to DLHSS (type 6, $\text{ITA} \leq 10^{\circ}$).}
\label{fig:EightDarkSamples}
\end{figure}

These eight samples are shown in row 1 of Figure~\ref{fig:EightDarkSamples} and indicate that images labelled as dark skin %are no images of dark skin, but 
are actually dark images of skin.
We qualitatively evaluated these images with a dermatologist who confirmed that they likely correspond to Fitzpatrick (FST) \cite{FitzpatrickVI} skin type III or lower.
%This illustrates the need to address problem \ref{ch:preprocessing}, to pre-process the images in colour space.
%
When RP2 labels images as dark skin and DLHSS does not, this can be explained by RP2 segmenting the lesion in all eight patches.
Since RP2 reports the brightest patch and DLHSS a median, DLHSS is likely more reliable in these cases.
On visual inspection of the five darkest images (DLHSS), this algorithm appears to be susceptible to hair and rare lesion sites, such as the tongue or ears; and dark skin labels can be explained by segmentation failures of DLHSS.
These results suggest that the ISIC18 dataset is not sufficiently diverse for a fairness analysis, as it presumably does not contain any images of dark skin (i.e. FST IV-VI \cite{FitzpatrickVI}).

\subsection{Fairness analysis}

\begin{figure}[h]
\centering
\scalebox{0.90}{
\centering
\includegraphics[width=\textwidth]{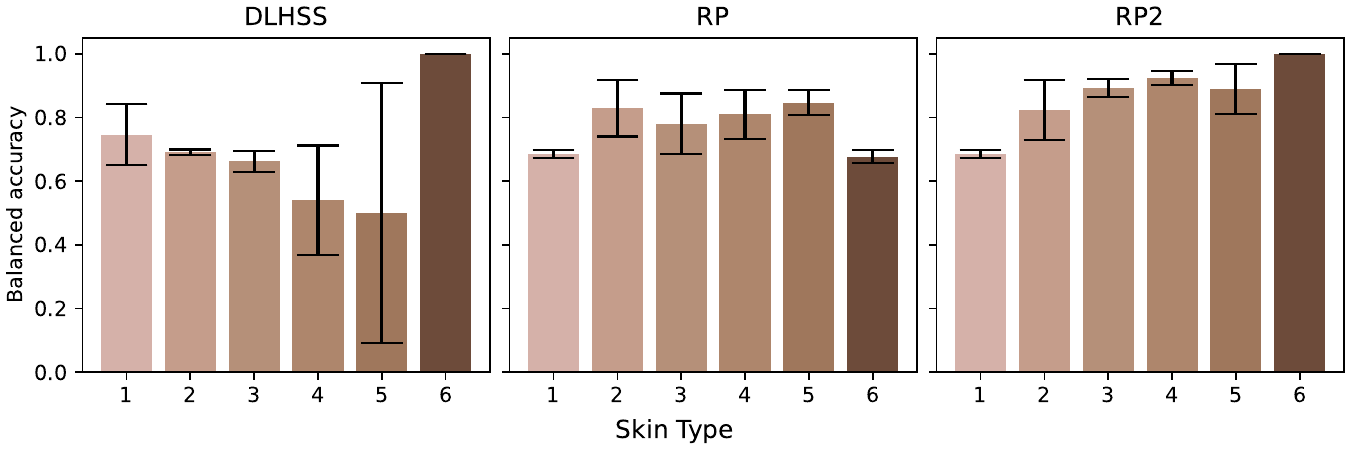}
}
\caption{Average balanced accuracy per skin type in the baseline experiment according to different automatic ITA estimation algorithms: Deep Learning-based Healthy Skin Segmentation (DLHSS), Random Patch algorithm with \textit{arctan} (RP) and with \textit{arctan2} (RP2). All bar charts stem from the same predictions.} \label{fig:FairnessBAcc}
\end{figure}

The question arises as to whether and how the different %automatic 
ITA estimation methods %described in Section~\ref{sec:AutoITAMethods}
impact the result of a downstream fairness analysis. % and potential mitigation techniques.
To this end, we analyse the balanced accuracy per skin type for the baseline experiment without data shift for DLHSS, RP, and RP2. %three ITA estimation algorithms are shown 
As shown in Figure~\ref{fig:FairnessBAcc}, we obtain different fairness results per method despite using the exact same classification model and test set.
%
%Although all three bar charts refer to the same predictions, they differ qualitatively.
For DLHSS ITA values, we note a decline in average balanced accuracy for types 1-4 (according to Figure \ref{subfig:ITAthresholds}).
%Note that there is a total of three and six Type 6 and Type 5 samples in the whole dataset, respectively, hence fewer samples in the test sets.
%Besides, the variance among different runs of the experiments is notable.
In contrast, the balanced accuracy of very light skin (type 1) and very dark skin (type 6) %samples 
according to RP is lower than for intermediate skin tones (types 2-5).
%, which is comparable to each other.
According to RP2, the light skin samples (type 1) show the worst performance, and the balanced accuracy appears to increase overall for darker skin types.
Thus, analysing the fairness of the same lesion predictions, the outcome changes depending on the applied ITA estimation method.

%\subsection{Data shift}
\subsection{Simulated data shifts} \label{sec:DataShiftExperiments}
\begin{figure}[tb]
\centering
\scalebox{0.90}{
\centering
\includegraphics[width=\textwidth]{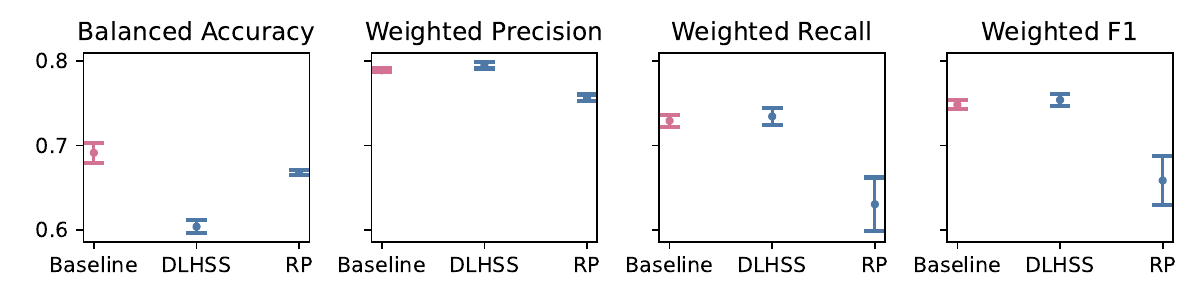}
}
\caption{Metrics for baseline and data shift experiments with the test set of the latter containing exclusively skin types 3-6 ($\text{ITA} \leq 41$ according to Figure \ref{subfig:ITAthresholds}). ITAs are obtained from DLHSS and RP estimation methods. Error bars show the standard deviation between three experiment repetitions with different random train-validation (DLHSS, RP) or train-validation-test (Baseline) splits.} \label{fig:DataShift}
\end{figure}

To simulate data shift, light skin images are used during training ($\text{ITA}>41^\circ$) while dark skin images are used in testing ($\text{ITA}\leq41^\circ$). %as in \cite{SkinToneBevan2022}
%, the test set is defined to exclusively contain dark skin of type 3 to 6 ($\text{ITA} \leq 41$, as in Figure \ref{subfig:ITAthresholds}). 
To assess whether the %choice of the automatic 
ITA estimation method impacts classification fairness %interpretation  of the classifier
in the presence of data shift, % experiments are conducted with
reproduced DLHSS and original RP ITA values are used.
%from the DLHSS algorithm \cite{IBM} and with the ITA labels reproduced from RP \cite{SkinToneBevan2022}.
%In the case of \cite{IBM}, there are five images, where an ITA estimate was not possible.
%These images are removed from the simulated data shift experiment using the ITA labels from \cite{IBM}, as they cannot be unambiguously mapped either to dark nor to light skin.
%As a baseline for comparison, a random split stratified by lesion type is defined as a control experiment.
%For comparison, a baseline is defined by randomly splitting the dataset stratified by lesion type while keeping the size of baseline and DLHSS training sets equal.
For comparison, the size of the baseline test set was defined equal to the DLHSS test set size and the train-validation ratio was 80:20 for all experiments.
%The training size in the baseline experiment is chosen equal to the training size of the data shift experiment using ITA estimations from \cite{IBM}, but with a stratified train-test split.
%The predictions of this baseline experiment are also assessed in terms of balanced accuracy per skin type like in \cite{IBM}, using the ITA labels from DLHSS \cite{IBM}, and the RP and RP2 algorithms.
%All experiments are repeated three times with different seeds to analyse the robustness.
%
The results %balanced accuracy and weighted precision, recall, and f1-score of the two data shift experiments and the baseline experiment 
of the data shift experiments are visualised in Figure \ref{fig:DataShift}. Despite a higher baseline accuracy, the weighted precision, recall, and f1-score, remain similar between the baseline and the data shift experiment with DLHSS labels. % does not appear to deteriorate classification performance.
%
%Since the test sets were defined under different criteria, the lesion occurrences need to be considered when interpreting Figure~\ref{fig:DataShift}, which are reported in Table \ref{tab:lesions}.
%
%\begin{table}[t]
%\centering \caption{Lesion type distributions in the ISIC18 dataset and the test sets for each experimental setup. The RP2 data shift experiment was discarded as the test was too unbalanced.}\label{tab:lesions}
%\centering
%\begin{tabular}{|l|r|r|r|r|r|r|r|r|}
%\hline
%{Lesion} & \textbf{AKIEC} & \textbf{BCC} & \textbf{BKL} & \textbf{DF} & \textbf{MEL} & \textbf{NV} & \textbf{VASC} & \textbf{TOTAL} \\
%\hline
%ISIC18 & 327 & 514 & 1,099 & 115 & 1,113 & 6,705 & 402 & 10,015\\
%Baseline & 96 & 150 & 321 & 33 & 325 & 1,960 & 42 & 2,927\\
%DLHSS & 89 & 81 & 412 & 22 & 310 & 1,981 & 32 & 2,927 \\
%RP & 32 & 105 & 302 & 27 & 297 & 1,509 & 29 & 2,301\\
%RP2 & 0 & 1 & 5 & 0 & 1 & 237 & 1 & 245 \\
%\hline
%\end{tabular}
%\end{table}
%
%The baseline experiment has a higher balanced accuracy than any data shift experiment.
%However, for the weighted precision, recall, and f1-score, the data shift with DLHSS labels does not appear to deteriorate classification performance.
The balanced accuracy is the \textit{unweighted} macro-averaged recall, hence the difference in balanced accuracy and \textit{weighted} (macro-averaged) recall suggests that the classification performance differs per lesion type.
%Noteworthy, 
This difference appears to depend on the train-test split induced by the ITA estimation method, yielding different lesion distributions.
Thus, not only the choice of evaluation metric, 
and the lesion type distribution,
but also the ITA estimations can alter the conclusions of the fairness analysis.

\section{Conclusions}
% 
%%%%%% (1) What we did
% Addressing fairness in automatic dermatological cancer diagnosis models is crucial due to variations in how skin diseases manifest across skin tones.
% Analysing fairness across skin types in deep learning dermatology, we compare skin tone estimation methods based on the Individual Typology Angle (ITA). 
%The skin tone labels have been estimated prior to fairness analysis in independent studies using the . 
%However, this work showed that although the distributions of skin tone estimates might appear similar, different ITA estimates do not necessarily refer to the same images as dark.
%%%%%% (2) ITA estimation very susceptible to preprocess (e.g. lighting), artifacts (e.g. hair), healthy skin definition (e.g. lesion contrast), pixel to image ITA   
We compared ITA-based skin tone estimation methods that revealed common pitfalls. Namely, susceptibility to lighting conditions, colour space calibration, presence of hair or dark edges. Moreover, we showed the effects of the differences in extracting the healthy skin, and differences in mapping ITA values from pixel to image-level. 
%%%%%% (3) Disagreement in ITA estimation
Further, we observed disagreements between ITA estimation methods, which did not necessarily refer to the same images as dark, and overestimation of dark samples in the ISIC18 dataset. 
%%%%%% (4) Overestimation of dark samples in datasets. 
%%%%%% (5) ISIC18 dataset [26] is not sufficiently diverse for a fairness analysis, as it does presumably not contain any images of dark skin (i.e. FST IV-VI) 
A qualitative analysis with a dermatologist revealed that the images, where different ITA estimation methods agree on their dark skin tones ($\text{ITA} \leq 28^{\circ}$), do not represent moderate brown to black skin types (FST IV-VI).
%%%%%% (6) ITA estimation algorithms impact the result of downstream fairness analysis
Furthermore, our skin lesion classification experiments demonstrated that the choice of ITA estimation method substantially impacts the results of classification fairness analyses.
%%%%%% (7) In data shift experiments, we show that not only the choice of evaluation metric, and the lesion type distribution, but also the ITA estimations can alter the conclusions of the fairness analysis
Our data shift experiments further confirmed that ITA estimation altered conclusions drawn from the fairness analysis.
% showed that not only the choice of the evaluation metric and the lesion type distribution, but also the ITA estimation alters conclusions drawn from the fairness analysis.
%%%%%% (8) Future work to improve upon results
We illustrated the need for more diverse dermatological datasets with diligently annotated skin tones, lighting conditions, dermatoscope and camera information.  %
Apart from improving ITA estimation, further avenues of research include the measurement of the differences in model calibration and in epistemic and aleatoric uncertainty per skin tone type. In the presence of limited datasets, synthetic samples with controllable skin tones and lesion types can be explored to quantify their effect on fairness.   
Overall, our work shows that current skin tone fairness assessments are inconclusive and further research is needed into unified and robust algorithms for automatic skin tone estimation to avoid carrying unwanted biases into dermatology practice when deploying deep learning models.
%The comparison of the random patch algorithms illustrates the importance of calibration in colour space before applying the ITA, which requires information about the illumination.
%With a myriad of dermatoscopes, cameras, and unknown lighting conditions, this task is far from trivial.
%We recommend addressing the four challenges -- pre-processing, skin segmentation, lesion segmentation, and a robust and meaningful ITA assignment.
%The limitations of current skin tone estimation methods might impact downstream bias mitigation techniques.
%In particular, we recommend a re-assessment of the experiments in \cite{SkinToneBevan2022} with ITA labels from the RP2 algorithm, as the majority of dark images among all skin types consisted of particularly pale images.
%Furthermore, due to the shortage of dark skin images in the ISIC18 dataset \cite{ISIC18}, we recommend to assess skin tone fairness on a different dataset in the future.

%Here an interesting paper to discuss for potential recommendation (e.g., robustness --- also across skin tones --- might be increased by acquisition of multiple images per lesion for training): \href{https://arxiv.org/abs/2306.02800}{paper} \\

%Here another paper for recommendation discussion: Testing dermatology classifiers on out-of-distribution data: \href{https://www.sciencedirect.com/science/article/pii/S0959804921004421}{paper}

%
% ---- Acknowledgements ----
%

\subsubsection*{Acknowledgements.}

This study has received funding from the European Union’s Horizon 2020 research and innovation programme under grant agreement No 952103. It was further partially supported by the project FUTURE-ES (PID2021-126724OB-I00) 
and by grant FJC2021-047659-I 
from the Ministry of Science and Innovation of Spain. 
%Kaisar Kushibar holds the Juan de la Cierva fellowship from the Ministry of Science and Innovation of Spain with a reference number FJC2021-047659-I.
We would like to thank Dr. Mireia Sábat (Hospital Parc Taulí) and Professor Rafael Garcia (Universitat de Girona) for interesting discussions on this work. 

\bibliographystyle{splncs04}
\bibliography{temp.bib}
%Author, F.: Article title. Journal %\textbf{2}(5), 99--110 (2016)

%Author, F., Author, S.: Title of a proceedings paper. In: Editor,
%F., Editor, S. (eds.) CONFERENCE 2016, LNCS, vol. 9999, pp. 1--13.
%Springer, Heidelberg (2016). \doi{10.10007/1234567890}

%Author, F., Author, S., Author, T.: Book title. 2nd edn. Publisher,

%Author, A.-B.: Contribution title. In: 9th International Proceedings
%on Proceedings, pp. 1--2. Publisher, Location (2010)

%LNCS Homepage, \url{http://www.springer.com/lncs}. Last accessed 4
%\end{thebibliography}
\end{document}